\documentclass[a4paper,fleqn,usenatbib]{mnras}     

\usepackage{ae,aecompl}

\usepackage{graphicx}   
\usepackage{amsmath}    
\usepackage{amssymb}    
\usepackage{pdflscape}
\usepackage{float}

\def\braket#1{\left<#1\right>}

\title[asteroseismic analysis of open cluster NGC 6811]{Asteroseismic analysis of eight solar-like oscillating evolved stars in the open cluster NGC 6811}

\author[\c{C}elik Orhan, Z.]{
Zeynep \c{C}elik Orhan$^{1}$\thanks{E-mail: zeynep.celik@ege.edu.tr}
\\
$^{1}$Department of Astronomy and Space Sciences, Science Faculty, Ege University, 35100, Bornova, \.{I}zmir, Turkey\\
}

\date{Accepted XXX. Received YYY; in original form ZZZ}

\pubyear{2021}

\begin{document}
\label{firstpage}
\pagerange{\pageref{firstpage}--\pageref{lastpage}}
\maketitle

\begin{abstract}
The {\emph{Kepler}} space telescope has provided exquisite data to perform asteroseismic analysis on evolved star ensembles. 
Studying star clusters offers significant insight into 
stellar evolution and structure, due to 
having a large number of 
stars with essentially the 
same age, distance, and chemical composition.
This study analysed eight solar-like oscillating evolved stars 
that are members of the open cluster NGC 6811 and 
modelled them for the first time.
The fundamental stellar parameters
are obtained from the interior model using
observational asteroseismic and non-asteroseismic constraints.
The stellar interior models are constructed using 
the {\small {MESA}} evolution code.
The mass-loss method is included in the 
interior models of the stars.
The stellar  masses and radii ranges of the stars 
are 2.23-2.40 $M_{\odot}$ and 8.47-12.38 $R_{\odot}$, respectively.
 Typical uncertainties for the mass and radius are $\sim$ 0.11  $M_{\odot}$ and $\sim$ 0.09 $R_{\odot}$, respectively.
The model masses and radii are compared with masses and radii obtained from asteroseismic and non-asteroseismic methods (scaling relations and classic methods).
The stellar ages fell in the range between $0.71$ and $0.82$ Gyr, with a typical uncertainty of $\sim$ $18$  per cent.
The model ages of the star calculated in this study 
are compatible with those reported in the literature for NGC 6811. 
\end{abstract}




\section{Introduction}

Asteroseismology
is a prominent observation method
that offers the potential to probe 
stellar interiors and test interior models. 
The best-fitting model 
and observed individual oscillation frequencies 
can give precise estimates of stellar parameters, 
but these parameters are model-dependent.
For example, the helium abundance is 
 an uncertain model input parameter.
It is difficult to obtain
the helium abundance
 from spectra,
thus
the approximate values given by the
Galactic element enrichment law is 
widely accepted in theoretical models.
Recent studies showed that 
analysing acoustic glitches in oscillation frequencies 
could constrain 
the surface helium abundances ($Y$), 
but the analysis requires very 
high-quality observed asteroseismic data
(Verma et al. 2014, 2017).
In addition, evolved stars are expected to 
lose mass on the red giant branch (RGB), 
but
the amount of mass loss is an important unsolved problem. 
Therefore, particular attention has been 
devoted to the three open clusters in the {\emph{Kepler}} field (NGC 6791, NGC 6811, and NGC 6819; Stello et al. 2010; Basu et al. 2011; Hekker et al. 2011; Stello et al. 2011a, 2011b; Miglio et al. 2012) due to the well-known fact that 
cluster stars share the same age, distance, 
and chemical composition, 
thereby allowing 
more stringent investigations into stellar evolution theory.
In this study, the interior models for eight evolved
stars in the NGC 6811
 are constructed 
using the {\small {MESA}} evolution code. 
By combining asteroseismic and non-asteroseismic 
observational constraints, fundamental stellar parameters
are obtained 
very precisely from interior models. 

The stellar mass plays the 
most important role in aiding further
understanding stellar evolution and structure,
therefore it needs to be determined very precisely.
However, determining the stellar masses from observations
is difficult, except for eclipsing binary stars.
Currently, the masses and radii of solar-like oscillating
stars can be directly determined 
using observational data.
When the large separation oscillation frequencies ($\Delta \nu$), 
frequency of the maximum amplitude ($\nu _{\rm max}$), 
and effective temperature ($T_{\rm eff}$) are known,
the masses and radii of solar-like oscillating stars
 are calculated from conventional scaling relations. 
These conventional scaling relations (Kjeldsen and Bedding 1995) assume that
the solar-like oscillating stars have similar internal structure conditions.
This assumption causes 
systematic errors 
in the $M$ and $R$ calculations.
A large number of studies 
have proposed new methods
of reducing these 
systematic errors 
(Bellinger 2019, Sharma et al. 2016, Y{\i}ld{\i}z, \c{C}elik Orhan \& Kayhan 2016; hereafter Paper III,
White et al. 2011; Mosser et al. 2013; Guggenberger et al. 2016, Y{\i}ld{\i}z et al. 2014a; hereafter Paper I; Y{\i}ld{\i}z et al. 2015; hereafter Paper II).

On the other hand, with the classical method, 
$M$ and $R$ are determined 
using non-asteroseismic observation parameters.
This method involves using 
distance to calculate stellar $M$ and $R$.
Thanks to the {\emph{Gaia}} space telescope, the distances of {\emph{Kepler}}
target stars are determined very precisely.
The distances values of eight evolved stars
 were taken from
 {\emph{Gaia}} database. 
In this method, 
the star's bolometric correction 
values from the
MESA Isochrones $\&$ Stellar Tracks
 ({\emph{MIST}}) bolometric tables are used. 
From here, the luminosity of the star is calculated. 
The stellar radius is determined from the luminosity. 
The mass is then calculated using the observational $log g$.
In this way, the $M$ and $R$ of the stars are determined
using {\emph{Gaia}} DR3 parallax, observed effective temperature and observed surface gravity
 and {\emph{MIST} bolometric correction tables (Choi et al. 2016; Dotter 2016). 

Another method is based on 
calculating the mass and radius by 
fitting stellar internal structure models with 
the non-asteroseismic observation parameters of the star.
However, the uncertainties in  $M$ and $R$ 
determined by this method are often large. 
Therefore, in this study, 
non-asteroseismic  ([Fe/H], distance, $logg$,
 and $T_{\rm eff}$)
and the standard
 asteroseismic observation parameters
($\Delta \nu$, $\nu _{\rm max}$) and,
 two reference frequencies 
($\nu _{\rm min0}$ and $\nu _{\rm min1}$)
are used while fitting the 
internal structure model.
The quantities 
($\nu _{\rm min0}$ and $\nu _{\rm min1}$)
are determined from a $\Delta \nu$
versus $\nu$ graph and 
are the references frequencies 
corresponding to the highest and lowest values of the 
pairwise $\Delta \nu$, respectively (see figure 3 of Paper I).
In this way, the best-fitting model 
that represents the star can be constructed.

In this study, eight solar-like evolved stars in the
 open cluster NGC 6811 were selected and analysed 
them using the {\small {MESA}} evolution code.
All stellar models have included the mass-loss method.
This method was necessary to calibrate
the stellar ages of the cluster member stars.
Model $M$ and $R$ values obtained by different methods
 were compared in detail. 
These different methods include 
conventional scaling relations, 
modified scaling relations (in Paper III), Bellinger methods (Bellinger 2019),
 and classic methods.
The model oscillation frequencies are calculated using 
the {\small {ADIPLS}} package (Christensen-Dalsgaard 2008).
The reference frequencies ($\nu _{\rm mins}$),
 other asteroseismic ($\nu _{\rm max}$ and $\Delta \nu$),
and non-asteroseismic parameters
(effective temperature, surface gravity, and
metallicity) are used to calibrate
the interior models.
Thus,
 $M$, $R$,
luminosity ($L$), gravity ($logg$)
and age ($t$) of the stars are obtained very precisely
from the interior models using observed 
asteroseismic and non-asteroseismic
 parameters.

This work is organised as follows. 
Section 2 presents
the observational properties of eight solar-like oscillating  evolved stars of NGC 6811.
The properties of the {\small {MESA}} evolution code, 
model calculation method, and the different methods used to calculate mass and radius are described in Section 3.
In Section 4, the model and scaling relations results obtained for these stars are compared.
Finally, Section 5 presents the results.

\section{ASTEROSEISMIC AND NON-ASTEROSEISMIC PROPERTIES OF THE eight evolved stars in NGC 6811}

\begin{figure}
\begin{center}
\includegraphics[width=\columnwidth]{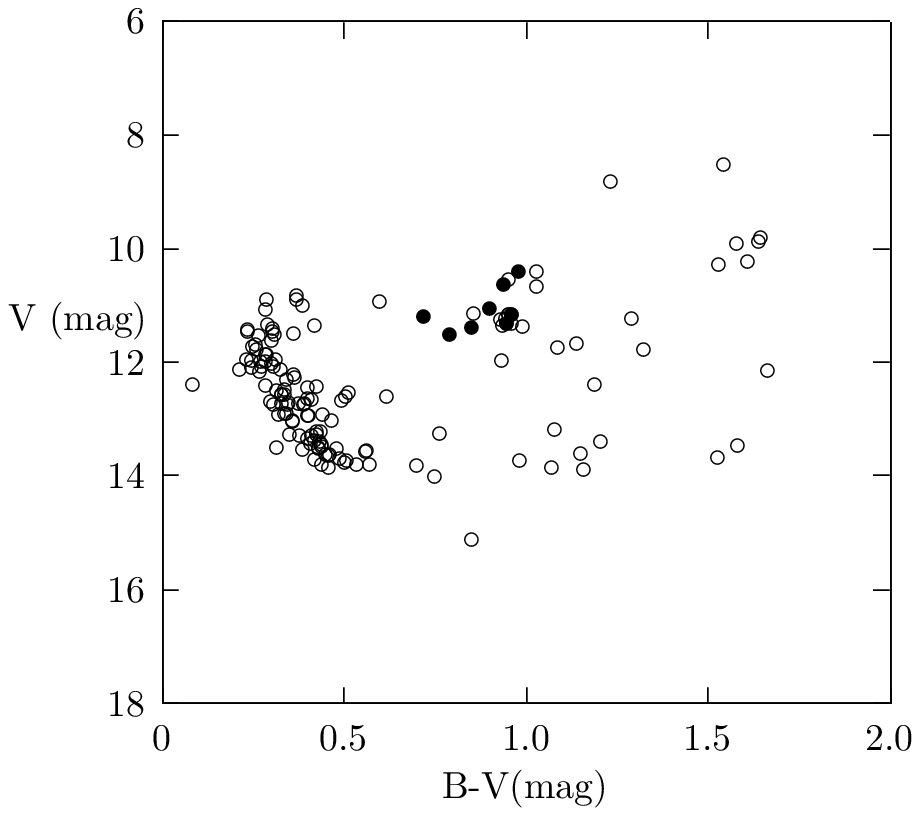}
        \caption{The B-V versus V diagram is plotted for eight solar-like oscillating evolved stars (filled circles) and other cluster member stars (circles) in NGC 6811.
}
\end{center}
\end{figure}


The period spacing carries 
information about the evolutionary state of 
an evolved star.
The range of observed period spacing of 
the stars is 100-190 s (Arentoft et al. 2017). 
As a result, when the observed period spacing 
values analysed using data from 
 Bedding et al. (2011, see fig.3), 
it shows that the evolved stars are
red-clump (RC)
(Pinsonneault et al. 2018).
According to the \emph{WEBDA} database,
the open cluster NGC 6811 is an intermediate-age ($log(age)=8.799$),
and slightly reddened ($E(B- V)=0.160$ mag) cluster
that is located at a distance of 1215 pc.
In Fig. 1, the eight evolved stars 
are plotted in a color-magnitude diagram.
While the filled circles in Fig. 1 represent 
the eight solar-like oscillating evolved stars modelled 
in this study, 
the circles represent the NGC 6811 cluster stars 
taken from \emph{WEBDA} database. 
Their B-V and V values were taken from \emph{SIMBAD} database.


The solar-like oscillations for 
the eight evolved stars in NGC 6811
have been detected 
by Arentoft et al. (2017). 
Table 1 lists the observed 
stellar asteroseismic and non-asteroseismic properties.
Their light curves are observed by 
the \textit{Kepler} space telescope. 
They show mixed mode oscillations on the observed frequencies. 
$\Delta\nu$, $\nu_{\rm \max}$, and mean small 
separation between oscillation frequencies 
(${\delta\nu_{\rm 02}}$) are derived by Arentoft et al. (2017). 
The observed and model  $\Delta\nu$
fit are determined from $\Delta\nu$ versus $\nu$ graphs (see Fig. 2).
Reference frequencies are also 
obtained from the $\Delta\nu$ versus $\nu$ graphs.

\begin{table*}
        \caption{Observed properties of eight solar-like oscillating evolved stars. The listed asteroseismic and non-asteroseismic properties are parallax ($\pi$) from {\emph{Gaia}} database and magnitude (G), effective temperature from spectra ($T_{\rm es}$), logarithmic of surface gravity ($log g$), $[Fe/H]$, ${\Delta\nu}$, small separation (${\delta\nu_{\rm 02}}$), ${\nu_{\rm max}}$, ${\nu_{\rm min0}}$, and ${\nu_{\rm min1}}$ respectively.}

        \begin{tabular}{lccccccccccccccccccccccrrr}
                \hline
KIC  &$\pi$&G&$T_{\rm es}$&$\log g$& [Fe/H] &${\Delta\nu}$ &${\nu_{\rm max}}$ & ${\nu_{\rm min0}}$&${\nu_{\rm min1}}$ \\ \\[1.2pt]
              &(mas)&(mag)&(K)&(cgs)&(dex)&(${\rm \mu}$Hz)&(${\rm \mu}$Hz)&(${\rm \mu}$Hz)&(${\rm \mu}$Hz)\\ \\[1.2pt]
  \hline

9409513&0.9866$\pm$0.02&10.24$\pm$0.03&4950$\pm$100&2.76$\pm$0.02&-0.050$\pm$0.050&6.04$\pm$0.02&69.8$\pm$1.0&81.75$\pm$3.04&61.90$\pm$3.04\\[2.0pt]
9532903&0.8485$\pm$0.02&10.91$\pm$0.03&5055$\pm$100&2.93$\pm$0.02&-0.007$\pm$0.020&7.55$\pm$0.04&92.0$\pm$1.5&105.86$\pm$3.78&81.05$\pm$3.78\\[2.0pt]
9534041&0.8516$\pm$0.02&11.02$\pm$0.03&5039$\pm$100&2.99$\pm$0.02&-0.052$\pm$0.050&8.35$\pm$0.01&103.8$\pm$1.0&117.64$\pm$4.18&87.08$\pm$4.18\\[2.0pt]
9655101&0.8658$\pm$0.02&10.97$\pm$0.03&5062$\pm$100&2.94$\pm$0.02&-0.039$\pm$0.050&7.88$\pm$0.02&98.7$\pm$1.0&103.50$\pm$3.94&83.89$\pm$3.94\\[2.0pt]
9655167&0.8652$\pm$0.02&11.04$\pm$0.03&5048$\pm$100&2.93$\pm$0.02&-0.006$\pm$0.020&8.07$\pm$0.01&99.4$\pm$2.1&105.23$\pm$4.04&88.99$\pm$4.04\\[2.0pt]
9716090&0.8483$\pm$0.02&11.09$\pm$0.03&5013$\pm$100&2.70$\pm$0.02&-0.050$\pm$0.050&8.54$\pm$0.02&107.8$\pm$1.4&119.78$\pm$4.27&94.11$\pm$4.27\\[2.0pt]
9716522&0.8124$\pm$0.02&10.41$\pm$0.05&4861$\pm$100&2.64$\pm$0.03&-0.030$\pm$0.020&4.88$\pm$0.01&53.7$\pm$1.0&58.10$\pm$2.44&48.21$\pm$2.44\\[2.0pt]
9776739&0.8639$\pm$0.02&10.90$\pm$0.05&5118$\pm$100&2.92$\pm$0.03&0.010$\pm$0.020&7.65$\pm$0.03&92.9$\pm$1.0&110.33$\pm$3.83&82.29$\pm$3.83\\[2.0pt]
               \hline
        \end{tabular}
\end{table*}

$\nu_{\rm \min0}$ and $\nu_{\rm \min1}$ have been determined 
from their $\Delta\nu$ versus $\nu$ graph
 for eight evolved stars (see Fig. 2). 
 As seen in Figure 2, decreases in 
observation and model frequencies are seen in 
the $\Delta\nu$ versus $\nu$ graph. The frequencies corresponding 
to the high and low frequencies from these drops
 are named as $\nu_{\rm \min1}$ and 
$\nu_{\rm \min0}$, respectively.
The value of $\nu_{\rm \min2}$ for these stars could not 
be determined from observed radial oscillation frequencies.
The method suggested in Paper I 
was used when determining these frequencies, which occur 
due to He II glitch.
First, the
frequency interval of the 
minimum in 
 $\Delta\nu$ versus $\nu$ graph was determined. 
In this frequency interval, two lines were 
drawn from the neighboring intervals.
The intersection of the two lines 
gives $\nu_{\rm \min}$ which are the reference frequencies.
Uncertainties in reference frequencies are 
calculated from half of the observed large separation (Paper III).
The uncertainties in $\nu_{\rm \min0}$ and $\nu_{\rm \min1}$ 
are listed in Table 1.

Spectroscopic data of 
 {\emph{Kepler}} target NGC 6811 cluster member stars 
($log g$, $[Fe/H]$, and $T_{\rm eff}$) are taken from Hawkins et al. (2016).
 Distance is a crucial stellar parameter. 
Thanks to the {\emph{Gaia}} space telescope, 
a large number of star distances are very precisely determined.
Parallax ($\pi$) and G magnitude are derived from the \emph{Gaia} DR3 database.
The effective temperature ranges between 4826 K (KIC 9716522) and 5027 K (KIC 9534041). 
The [Fe/H] ranges between 
-0.052 (KIC 9534041) and 0.010 dex (KIC 9776739).

\section{Properties of Interior models of the stars}
\subsection{Properties of {\small {MESA}} evolution code}
The stellar interior models are constructed using 
the {\small {MESA}} evolution code 
(version 15140, Paxton et al. 2011, 2013). 
Based on calibration of the solar model,
the values Y, Z and, 
convective parameters ($\alpha$) are computed as
0.2745, 0.0172, and 1.8125, respectively.

Convection is treated with a standard mixing-length theory (B\"{o}hm- Vitense 1958).
The interior models do not consider
microscopic diffusion effects.
A small amount of convective overshoot as
described by Herwig (2000) was allowed
during both the main-sequence
and red giant phases.
The exponential convective overshoot
parameter was fixed at $f _{\rm 0}$=0.016, 
based on the values used in the MIST isochrones (Choi et al. 2016).
 {\small {MESA}} offers the opacity tables 
of Iglesias \& Rogers (1993, 1996) and
includes their OPAL opacity tables 
in the high-temperature region supplemented 
by low-temperature 
tables of Ferguson et al. (2005) 
with fixed metallicity as the a default option.

Stellar metallicity significantly affects 
pre-main sequence (pre-MS) phases as 
well as every other 
phase of the star. 
 It is therefore important for us 
to start from this phase 
and evolve the star 
there.
In this study, 
the pre-MS is included in the 
construction of stellar interior models. 
Nuclear reaction rates are taken 
from Angulo et al. (1999) and Caughlan \& Fowler (1988).
Mass-loss can be included into {\small {MESA}} in a variety of built-in parametrizations. 
Reimers' mass-loss law is applied for the stellar evolution models.
Atmospheric conditions significantly effect the
model oscillation frequencies determined.
The \texttt{simple$\_$photosphere} option
is selected in the {\small {MESA}}
code for the star interior models.
The theoretical oscillation frequencies 
are computed by 
the {\small {ADIPLS}} pulsation package (Christensen-Dalsgaard 2008).
To correct the model frequencies for so-called 
"surface effects" due to incomplete 
modeling of the near-surface layers, 
the empirical prescription of Kjeldsen et al. (2008) is used.

\begin{table*}
 \caption{List of model properties of evolved solar-like oscillating stars.
$M_{\rm mod}$, $R_{\rm mod}$, ${T_{\rm mod}}$, $L_{\rm mod}$, ${\log g_{\rm mod}}$, $\alpha$, $t_{\rm mod}$, and ${{\chi^2_{\rm spec}}}$,  respectively, stellar mass in $M_{\sun}$ unit, stellar radius in $R_{\sun}$ unit, effective temperature in $K$ unit, luminosity in $L_{\sun}$ unit, logarithm of surface gravity of the model and stellar age in Gyr unit.}

         \begin{tabular}{lcccccccccr}

       \hline
        KIC   & $M_{\rm mod}$  & $R_{\rm mod}$ & $T_{\rm mod}$ & $L_{\rm mod}$ & ${\log g_{\rm mod}}$&  $\alpha$& $t_{\rm mod}$ & ${{\chi^2_{\rm spec}}}$ \\
              & $(M_{\sun})$ &$(R_{\sun})$ & (K)  & $(L_{\sun})$ &(cgs)&  &(Gyr)&     \\
                \hline
9409513&2.40$\pm$0.10&10.69$\pm$0.09&4985$\pm$100&55.21$\pm$3.41&2.76$\pm$0.01&2.07&0.71$\pm$0.18&0.25\\[2pt]
9532903&2.27$\pm$0.11&9.17$\pm$0.08&4992$\pm$100&42.48$\pm$2.07&2.87$\pm$0.02&1.55&0.79$\pm$0.17&0.23\\[2pt]
9534041&2.38$\pm$0.10&9.05$\pm$0.08&5025$\pm$100&53.37$\pm$2.34&2.90$\pm$0.02&1.95&0.72$\pm$0.18&0.13\\[2pt]
9655101&2.35$\pm$0.11&9.09$\pm$0.09&5060$\pm$100&51.00$\pm$2.92&2.91$\pm$0.02&1.95&0.73$\pm$0.18&0.15\\[2pt]
9655167&2.25$\pm$0.11&8.79$\pm$0.09&5044$\pm$100&45.60$\pm$2.90&2.90$\pm$0.01&1.80&0.82$\pm$0.18&0.25\\[2pt]
9716090&2.23$\pm$0.12&8.47$\pm$0.08&4997$\pm$100&37.18$\pm$2.45&2.93$\pm$0.01&1.60&0.75$\pm$0.18&0.13\\[2pt]
9716522&2.35$\pm$0.12&12.38$\pm$0.12&4824$\pm$100&83.41$\pm$4.80&2.62$\pm$0.01&1.95&0.77$\pm$0.16&0.05\\[2pt]
9776739&2.34$\pm$0.14&9.24$\pm$0.09&5120$\pm$100&49.18$\pm$3.74&2.88$\pm$0.01&1.75&0.75$\pm$0.16&0.07\\[2pt]

               \hline
        \end{tabular}

\end{table*}

\subsubsection{Modelling strategy and uncertainty calculation}
The model input parameters are initial $M$, $Y$,
$Z$, and $\alpha$.
When a stellar model is constructed,
all available observational data for that
particular star are used.
$M$ is the most important parameter 
affecting stellar structure and evolution.
Therefore, an initial mass value must be 
entered in the model calculations.
This study uses three different methods
 to determine the initial mass value (see Sect.3.1).
Since scaling relations 
are developed for MS stars, 
$M_{\rm \pi}$ is taken as the first 
mass in the constructed star model.
To calibrate the stellar interior models, the initial values of $M$ 
and $\alpha$ were changed until
the observed asteroseismic and non-asteroseismic 
constraints were obtained
in the models.

\begin{figure}
\begin{center}
\includegraphics[width=\columnwidth]{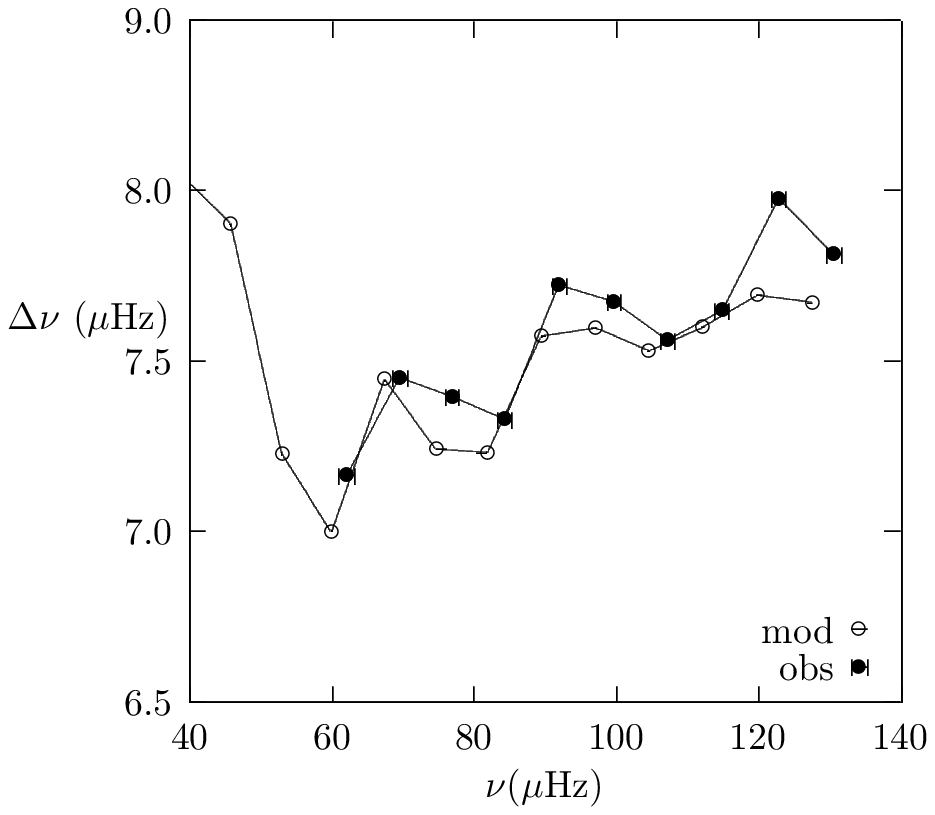}
        \caption{Plot of ${\Delta\nu}$ as a function of ${\nu}$ graph for model ({\small MESA}, circles) and observed frequencies (filled circles) of KIC 9776739.
}
\end{center}
\end{figure}

After mass, $Z$ is the parameter
that mostly affects the stellar structure and evolution.  
Therefore, $Z$ is another important input parameter. 
In general, $Z$ is computed from the observed [Fe/H] 
or overall metallicity [M/H] reported in the literature.
Then the total metallicity 
($Z$) from [M/H] is computed 
using this method. 
$Z=10^{\rm [Fe/H]} 0.0134$ 
is used for metallicity calculations of initial models.
The metallicity in the present-day solar photosphere 
was determined as
0.0134 by Asplund et al. (2019). 
Therefore, in this study, the solar metallicity value of 
the eight evolved stars is taken as 0.0134.
Here, the lowest and the highest
metal abundance are calculated as 0.0119 and 0.0370, respectively.
The average of the calculated metallicity of 
the stars is determined as 0.0126
and the initial
$Y$ is set to as 0.2745 (solar value).

The model calibration procedure 
is the same as that of \c{C}elik Orhan, Y{\i}ld{\i}z \& Kayhan (2021) 
for modelling of 15 evolved stars (see Section 3.2).
The compatibility between models and the 
non-asteroseismic and asteroseismic 
observational data is examined.
For this purpose ${{\chi^2_{\rm spec}}}$ values are calculated 
from Eq. 6 (see Sec. 3.3). 
While examining the harmony between the 
observed and model parameters, 
special attention was paid to
 make sure that the ${{\chi^2_{\rm spec}}}$ value is smaller than one. 
The obtained ${{\chi^2_{\rm spec}}}$ values are listed in Table 2.

In addition, reference frequencies 
and other seismic parameters were also examined,
 while seeking to determine the best model.
In Fig 2, ${\Delta\nu}$ versus ${\nu}$ 
for KIC 9776739 is plotted for comparing 
 observed and model oscillation frequencies 
as well as the reference frequencies.
The observation frequencies of 
each star are 
close to those computed by the models.
Special attention was paid to make 
sure that the model and 
observational ${\Delta\nu}$ values were the same. 
The fact that the patterns 
overlapped and 
the same on the  ${\Delta\nu}$ versus ${\nu}$ graph 
is the most important feature of the model oscillation frequencies.
Attention is paid to ensure that this value is below 
than one for the interior models.
The obtained ${{\chi^2_{\rm seis}}}$ are listed in Table 3.
\subsubsection{Mass loss in interior models}
RGB mass loss is the content of most stellar 
evolution codes, and is generally obtained by 
simple relations of the stellar parameters.
The most commonly used mass-loss method, 
the Reimers (1975) method, 
was used in this stellar interior models 
of these eight evolved stars in this study.
However, using this method requires a reference point ($\eta$).
Assuming a Reimers mass-loss law, 
model comparisons suggest $\eta$ in the range 
0-0.1 with a strong preference for no mass loss at all.

Red-clump (RC) stars are the only ones for which classical mass-loss
formulas (Reimers 1975, Renzini \& Fusi Pecci 1988)
can produce an appreciable mass loss.
There are eight RC stars in this study.
The stellar
interior models of RC stars are 
constructed using Reimers (1975) mass-loss methods ($\eta$=0.1). 
The total mass loss in all the internal structure 
models constructed is 0.01 $M_{\rm \odot}$.

\subsubsection{The uncertainties calculations}
To quantify the differences between 
the model and observational data 
obtained for each star, a normalised 
$\chi^2$ was calculated for 
asteroseismic and non-asteroseismic constraints:

\begin{equation}
{\chi^2_{\rm seis}}= \frac{1}{N_{\rm f}}\sum\limits_{i=1}^n\left(\frac{{\nu_{\rm i,obs}}-{\nu_{\rm i,mod}}}{\sigma_{\rm i,obs}}\right)^2
\end{equation}
and
\begin{equation}
{\chi^2_{\rm spec}}= \frac{1}{N_{\rm s}}\sum\limits_{i=1}^n\left(\frac{{P_{\rm i,obs}}-{P_{\rm i,mod}}}{\sigma_{\rm i,obs}}\right)^2
\end{equation}
where ${\nu_{\rm i,obs}}$ and, ${\nu_{\rm i,mod}}$ are the observed and model oscillation frequencies, respectively.
${N_{\rm f}}$ is the total number of the modes for $l=0$, ${\sigma_{\rm i,obs}}$ is
the  uncertainty of the observed frequencies.
${\nu_{\rm max}}$, ${\Delta\nu}$,
${\nu_{\rm min0}}$ and, ${\nu_{\rm min1}}$
 are used to calculate ${{\chi^2_{\rm seis}}}$.
The ${P_{\rm i,obs}}$ is
non-asteroseismic observed data 
(${T_{\rm eff}}$, $\log g$). 
${N_{\rm s}}$ total number of the data included
and ${P_{\rm i,mod}}$ are non-asteroseismic parameters of the models.

Monte Carlo simulations are used 
in the error calculations of $M$, $R$, $logg$, 
$L$, and $t$ values of the 
eight evolved stars determined by the model.
The uncertainties are listed in Table 2.
The uncertainty of ${T_{\rm mod}}$ is considered the same as ${\Delta T_{\rm es}}$.
 ${\Delta Y_{\rm mod}}$/ ${Y_{\rm mod}}$$\approx$$3{\Delta M_{\rm mod}}$/${M_{\rm mod}}$ is obtained. This expression is
derived from the interior model data. 
For the stars ${\Delta Z_{\rm mod}}$ is computed from the observed uncertainty of [Fe/H].
The mean ${\Delta Z_{\rm mod}}$ 
is determined by averaging all the 
calculated ${\Delta Z_{\rm mod}}$ values.
The values of these approaches
 are determined as Y=0.2745$\pm$0.0824 and Z=0.0126$\pm$0.0003 
with certain error margins.

\begin{table*}
        \caption{Asteroseismic parameters of {\small MESA} model results of the eight evolved stars.
                  ${\braket{\Delta{\nu_{\rm M}}}}$, ${\nu_{\rm max,M}}$,
                 ${\nu_{\rm min0,M}}$, ${\nu_{\rm min1,M}}$, and ${\nu_{\rm min2,M}}$ are, respectively, large separation between model oscillation frequencies,
                 model oscillation frequency of maximum amplitude, reference frequencies of model in $\mu$Hz units. ${\nu_{\rm max,M}}$ is computed from scaling relations with
                 ${T_{\rm eff,M}}$ and ${logg_{\rm M}}$. ${{\chi^2_{\rm seis}}}$ of the model is in the last column.Typical uncertainties for the reference frequencies are ${\braket{\Delta{\nu_{\rm M}}}}/2$. }
        \begin{tabular}{lcrrrrrr}
                \hline
         KIC   & ${\braket{\Delta{\nu_{\rm M}}}}$ & ${\nu_{\rm max,M}}$ & ${\nu_{\rm min0,M}}$ &
          ${\nu_{\rm min1,M}}$ & ${\nu_{\rm min2,M}}$& ${{\chi^2_{\rm seis}}}$ \\
         & ($\mu$Hz) & ($\mu$Hz) & ($\mu$Hz) & ($\mu$Hz) & ($\mu$Hz) &  \\
                \hline
9409513&6.02&68.21&79.86&59.04&53.42&0.69\\
9532903&7.53&93.17&106.83&81.19&61.65&0.45\\
9534041&8.32&114.11&83.86&62.285&61.66&0.47\\
9655101&7.88&96.55&103.45&83.74&64.22&0.32\\
9655167&8.04&98.256&107.23&85.84&64.71&0.65\\
9716090&8.49&116.176&119.45&92.14&69.07&0.43\\
9716522&4.87&56.159&57.53&47.45&33.74&0.45\\
9776739&7.62&93.168&104.04&77.93&62.70&0.48\\

                \hline
        \end{tabular}
\end{table*}

\subsection{Mass and radius calculation from scaling relations}
In this study, stellar $M$ and $R$ values are calculated based on three different scaling relations: relations are
conventional scaling relations, 
modified scaling relations, and 
the Bellinger method (Bellinger 2019).
However, conventional and  
modified scaling relations methods 
were derived for MS stars.
That's why 
the methods 
should be tested for evolved stars.
The conventional relations are (Kjeldsen \& Bedding 1995):
\begin{equation}
\frac{M_{\rm sca}}{M_{\rm \odot}}= \left(\frac{\nu_{\rm max}}{\nu_{\rm max\odot}}\right)^3\left(\frac{{{\Delta\nu_{\rm }}}}{{{\Delta\nu_{\rm \odot}}}}\right)^{-4}\left(\frac{T_{\rm eff}}{T_{\rm eff \odot}}\right)^{1.5}, 
\end{equation}
and
\begin{equation}
\frac{R_{\rm sca}}{R_{\rm \odot}}= \left(\frac{\nu_{\rm max}}{\nu_{\rm max\odot}}\right)\left(\frac{{{\Delta\nu_{\rm }}}}{{{\Delta\nu_{\rm \odot}}}}\right)^{-2}\left(\frac{T_{\rm eff}}{T_{\rm eff \odot}}\right)^{0.5}. 
\end{equation}
${\nu_{\rm max, \odot}}$  and  ${\Delta\nu_{\rm \odot}}$ are taken as ${\nu_{\rm max, \odot}}$=3050 ${\rm \mu}$Hz (Kjeldsen \& Bedding 1995) and ${\Delta\nu_{\rm \odot}}$=135.15 ${\rm \mu }$Hz from BiSON solar data (Chaplin et al. 2014).


\begin{table*}
        \caption{Listing of all obtained masses and radii of the eight evolved stars. ${M_{\rm sca}}$ and ${R_{\rm sca}}$ are calculated from conventional scaling relations. ${M_{\rm III}}$ and ${R_{\rm III}}$ are determined using formulas from Paper III. ${M_{\rm \pi}}$ and ${R_{\rm \pi}}$ are calculated from classic method. 
${M_{\rm bel}}$ and ${R_{\rm bel}}$ are calculated from Python code delivered by Bellinger (2019).  {${M_{\rm aren}}$ and ${R_{\rm aren}}$ are taken from Arentoft et al. (2017). }
{\small MESA} masses and radii are given as ${M_{\rm mod}}$ and ${R_{\rm mod}}$, respectively. 
}
\resizebox{\textwidth}{!}{%
   \begin{tabular}{lcrrrrrrrrrrrrrrrrr}
                \hline
         KIC   & $M_{\rm sca}$ & $R_{\rm sca}$ & $M_{\rm III}$ & $R_{\rm III}$&  $M_{\rm \pi}$ & $R_{\rm \pi}$& 
$M_{\rm bel}$&$R_{\rm bel}$&
$M_{\rm aren}$&$R_{\rm aren}$& 
           $M_{\rm mod}$ & $R_{\rm mod}$ &
                                         \\
         & ($M_{\sun}$)&($R_{\sun}$) &($M_{\sun}$) &($R_{\sun}$)&($M_{\sun}$) &($R_{\sun}$) &($M_{\sun}$) &($R_{\sun}$) 
&($M_{\sun}$) &($R_{\sun}$)
&$M_{\sun}$ &$R_{\sun}$  
                                             \\
                \hline
9409513&2.29$\pm$0.12&10.46$\pm$0.20&2.15$\pm$0.05&9.69$\pm$0.07&2.06$\pm$0.09&9.92$\pm$0.10&2.30$\pm$0.13&10.58$\pm$0.21&2.35$\pm$0.07&10.61$\pm$0.13&2.40$\pm$0.10&10.69$\pm$0.09\\[2pt]
9532903&2.15$\pm$0.12&8.83$\pm$0.17&2.35$\pm$0.07&9.08$\pm$0.07&2.28$\pm$0.07&9.05$\pm$0.07&2.17$\pm$0.13&8.94$\pm$0.20&2.21$\pm$0.07&9.00$\pm$0.13&2.27$\pm$0.11&8.92$\pm$0.08\\[2pt]
9534041&2.09$\pm$0.10&8.17$\pm$0.13&2.22$\pm$0.05&8.81$\pm$0.04&2.25$\pm$0.05&8.46$\pm$0.09&2.10$\pm$0.10&8.26$\pm$0.13&2.11$\pm$0.01&8.24$\pm$0.06&2.38$\pm$0.12&9.05$\pm$0.08\\[2pt]
9655101&2.24$\pm$0.11&8.70$\pm$0.14&2.45$\pm$0.03&8.97$\pm$0.03&2.11$\pm$0.03&8.41$\pm$0.03&2.25$\pm$0.10&8.79$\pm$0.15&2.23$\pm$0.09&8.75$\pm$0.70&2.35$\pm$0.11&9.09$\pm$0.09\\[2pt]
9655167&2.03$\pm$0.15&8.29$\pm$0.20&2.25$\pm$0.03&8.57$\pm$0.03&2.13$\pm$0.03&8.41$\pm$0.03&1.97$\pm$0.14&8.25$\pm$0.20&2.15$\pm$0.05&8.49$\pm$0.10&2.25$\pm$0.11&8.79$\pm$0.09\\[2pt]
9716090&2.10$\pm$0.13&8.17$\pm$0.13&2.23$\pm$0.03&8.35$\pm$0.03&2.19$\pm$0.03&8.58$\pm$0.03&2.11$\pm$0.11&8.15$\pm$0.15&2.10$\pm$0.00&8.07$\pm$0.00&2.23$\pm$0.12&8.47$\pm$0.08\\[2pt]
9716522&2.33$\pm$0.17&12.12$\pm$0.30&2.30$\pm$0.05&11.87$\pm$0.02&2.36$\pm$0.05&12.20$\pm$0.05&2.33$\pm$0.15&12.24$\pm$0.28&2.33$\pm$0.11&12.21$\pm$0.16&2.35$\pm$0.12&12.38$\pm$0.12\\[2pt]
9776739&2.07$\pm$0.15&8.64$\pm$0.21&2.41$\pm$0.05&9.14$\pm$0.05&2.18$\pm$0.05&8.77$\pm$0.05&2.08$\pm$0.10&8.73$\pm$0.16&2.21$\pm$0.06&8.91$\pm$0.16&2.34$\pm$0.14&9.24$\pm$0.09\\[2pt]
               \hline
        \end{tabular}
}
\end{table*}

Secondly, $M_{\rm III}$ and $R_{\rm III}$ are calculated from the modified scaling relation from Paper III. 
This work assumed that 
the first adiabatic exponent at the stellar surface 
(${\Gamma_{\rm 1s}}$) affects 
the relationship 
between ${\Delta\nu}$ and the square root of its mean density 
and developed a new relation between 
$\Delta\nu$, $\rho$, and ${\Gamma_{\rm 1s}}$.
The modified scaling relations are
as follows:

\begin{equation}
\frac{M_{\rm III}}{M_{\rm \odot}}= \left(\frac{\nu_{\rm max}}{\nu_{\rm max\odot}}\right)^3\left(\frac{{{\Delta\nu_{\rm }}}}{{{\Delta\nu_{\rm \odot}}}}\right)^{-4}\left(\frac{T_{\rm eff}}{T_{\rm eff \odot}}\frac{\Gamma_{\rm 1s \odot}}{\Gamma_{\rm 1s}}\right)^{1.5}\left(\frac{f^{4}_{\rm {\Delta\nu}}}{f^{3}_{\rm {\nu}}}\right)
\end{equation}
and

\begin{equation}
\frac{R_{\rm III}}{R_{\rm \odot}}= \left(\frac{\nu_{\rm max}}{\nu_{\rm max\odot}}\right)\left(\frac{{{\Delta\nu_{\rm }}}}{{{\Delta\nu_{\rm \odot}}}}\right)^{-2}\left(\frac{T_{\rm eff}}{T_{\rm eff \odot}}\frac{\Gamma_{\rm 1s \odot}}{\Gamma_{\rm 1s}}\right)^{0.5}\left(\frac{f^{2}_{\rm {\Delta\nu}}}{f_{\rm {\nu}}}\right).
\end{equation}

Thirdly, $M_{\rm bel}$ and $R_{\rm bel}$ are calculated from Python code derived from the Bellinger method (2019). 
In this code, the observed ${\nu_{\rm max}}$, ${\Delta\nu}$, ${T_{\rm eff}}$,
 and [Fe/H] values are entered together with errors.
In this way, values of  $M$, $R$, and $t$ are calculated from the 
scaling relations developed for evolved stars.
The $M$ and $R$ of the stars obtained by these relations are given in Table 4.

\subsection{Mass and radius calculation from MIST table}
The $M_{\rm \pi}$ and $R_{\rm \pi}$ are computed using the MIST bolometric correction table 
(Choi et al. 2016; Dotter 2016). 
Thus, $L$ values are determined for 
the stars for whiche the distances are known.
From this calculated $L$ values, the $R$ and $M$ values
 can be determined by the classical method.

While calculating $M_{\rm \pi}$ and $R_{\rm \pi}$ using 
this method, {\emph{Gaia}} DR3 distance and 
G magnitude, ${T_{\rm es}}$, and $log g$ input parameters are used. 
The trigonometric parallaxes from 
{\emph{Gaia}} DR3 are known to contain 
a zero-point offset that has a complex dependence 
on other observational parameters (e.g. the colour and magnitude of the stars).
In this study, the parallax 
offset is taken 0.028 mas (Huang et al. 2021).

First, the stellar $L$ is calculated 
for the stars using the MIST bolometric correction table.
R is obtained by the L=4$\pi$$R^2$$\sigma$${T_{\rm eff}}^4$ equation. 
After determining the radius, the $M$ is calculated from the $g=GM/R^2$ relation.
The precision of determining $M$ by this method depends on the sensitivity of the $log g$ value derived from the observations.
For this reason, the observed $log g$ 
value determined from the asteroseismic data is preferred, 
when obtaining the $M_{\rm \pi}$ and $R_{\rm \pi}$ of the stars. 
$M_{\rm \pi}$ and $R_{\rm \pi}$ values
calculated by the classical method are listed in Table 4.

\section{Results}
\begin{figure}
\begin{center}
\includegraphics[width=\columnwidth]{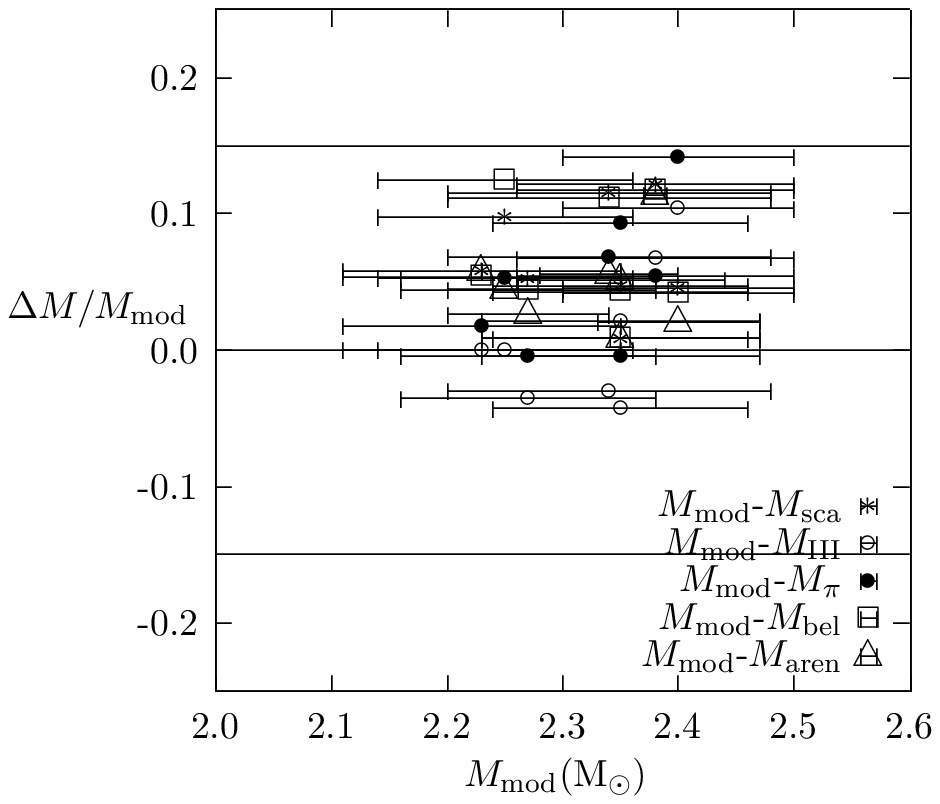}
        \caption{The mass differences are plotted for eight solar-like oscillating evolved star in the NGC 6811. The asterisks, circles, filled circles, squares, and triangles indicate ${M_{\rm mod}}$-${M_{\rm sca}}$, ${M_{\rm mod}}$-${M_{\rm III}}$, ${M_{\rm mod}}$-${M_{\rm \pi}}$, ${M_{\rm mod}}$-${M_{\rm bel}}$, and ${M_{\rm mod}}$-${M_{\rm aren}}$ respectively. The lines for -0.15 and 0.15.
}
\end{center}
\end{figure}
\subsection{Mass and radius comparisons}
Determining the stellar masses by direct observations is difficult, except for eclipsing binary stars. 
However, stellar $M$ can be determined using 
observational data with scaling relations
 developed with asteroseismology.
In this study, $M_{\rm sca}$, $M_{\rm III}$ and $M_{\rm bel}$ are calculated from 
conventional, modified scaling relations, and Bellinger method, respectively (see Sec. 3.2).
In addition, 
the masses of those stars for 
which distances are known are calculated 
by the classical method ($M_{\rm \pi}$) (see Sec. 3.3).
Lastly, $M_{\rm aren}$ is taken from Arentoft et al. (2017).
The $M_{\rm aren}$ is obtained using BaSTI grid modeling method.
They fit the observed asteroseismic and 
atmospheric quantities to a grid of BaSTI (Pietrinferni et al. 2004) isochrones 
using the Bayesian Stellar Algorithm (Silva Aguirre et al. 2015, 2017).  
The masses calculated using these different 
methods are compared with those determined from the models ($M_{\rm mod}$).
In Fig. 3, the asterisks, circles, filled circles, squares, and triangles indicate the $M_{\rm mod}$-$M_{\rm sca}$, $M_{\rm mod}$-$M_{\rm III}$, 
$M_{\rm mod}$-$M_{\rm bel}$,
 $M_{\rm mod}$-$M_{\rm \pi}$,
 and $M_{\rm mod}$-$M_{\rm aren}$  respectively. 
In Fig 3, in general, the difference between model and calculated $M$ values are less than  1.5 $\%$.
The model mass of KIC 9655101 is the same as $M_{\rm III}$.
The harmony between the $M_{\rm mod}$ and $M_{\rm \pi}$ is shown in Fig. 3.
All $M$ values are listed in Table 4.

The observed oscillation frequencies, 
particularly ${\Delta\nu}$,
 are related 
on the stellar $R$, because
the observed ${\Delta\nu}$ in a star 
depends on the mean density 
and the mean density is inversely 
proportional to the radius.
 If the observed oscillation frequencies are fitted 
from the model oscillation frequencies, the stellar $R$ is determined very precisely from interior models.
In Fig. 4, the radii obtained 
from different asteroseismic methods and 
the classic methods were compared to 
the radius obtained from the model.
The stellar $R$ values calculated 
from different methods 
are in good agreement with $R_{\rm mod}$.
$R$ values are listed in Table 4.
The difference between the 
$R_{\rm mod}$ and  $R_{\rm sca}$, $R_{\rm III}$,  $R_{\rm bel}$, and 
$R_{\rm aren}$ is 1.0 $\%$.
In addition, the difference between $R_{\rm \pi}$ and $R_{\rm mod}$ is about less than 1.0 $\%$. 
  

\begin{figure}
\begin{center}
\includegraphics[width=\columnwidth]{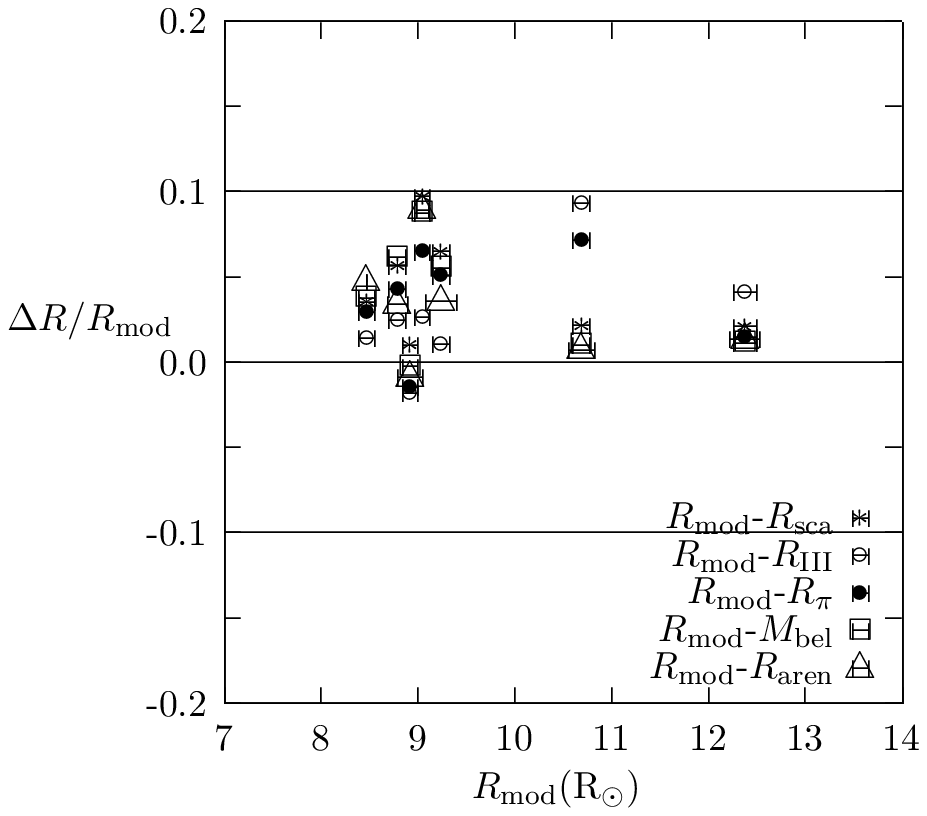}
        \caption{The stellar radii differences are plotted for eight solar-like oscillating stars in the  NGC 6811. The asterisks, circles, filled circles, squares, and triangles indicate ${R_{\rm mod}}$-${R_{\rm sca}}$, ${R_{\rm mod}}$-${R_{\rm III}}$, ${R_{\rm mod}}$-${R_{\rm \pi}}$, ${R_{\rm mod}}$-${R_{\rm bel}}$, and ${R_{\rm mod}}$-${R_{\rm aren}}$ respectively. The lines are for -0.10 and 0.10.}
\end{center}
\end{figure}

\begin{figure}
\begin{center}
\includegraphics[width=\columnwidth]{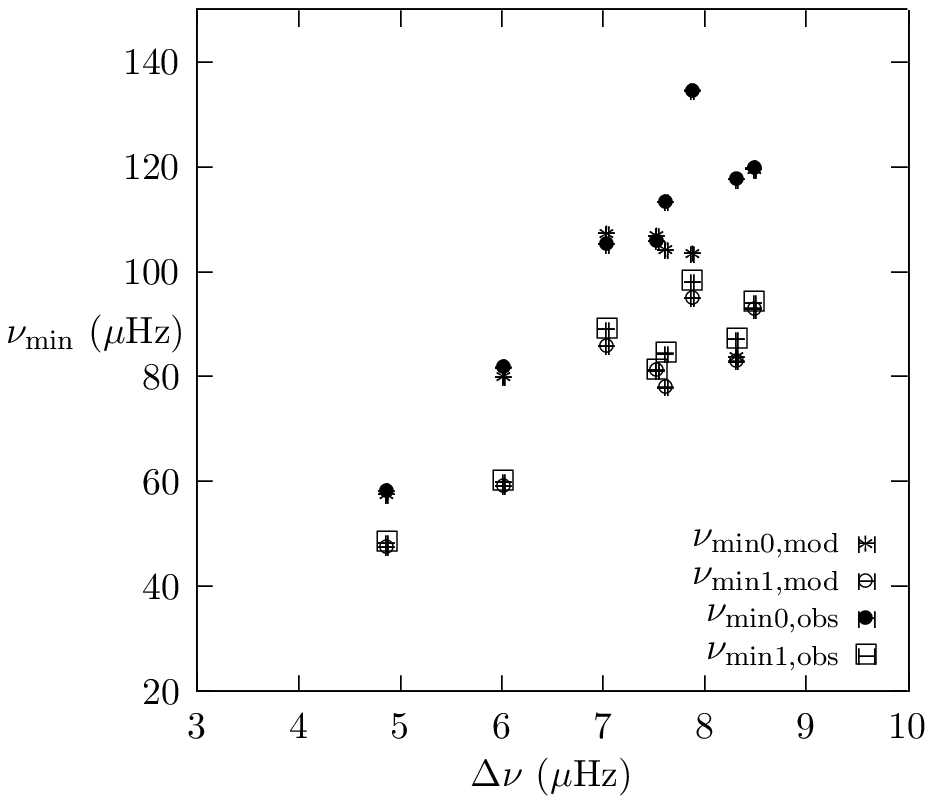}
        \caption{Plot of $\Delta\nu$ as a function of the  reference frequencies (${{\nu_{\rm min0}}}$ and ${{\nu_{\rm min1}}}$) for the stars. The asterisks and filled circles are the represented values of ${{\nu_{\rm min0}}}$ obtained from model and observed oscillation frequencies, respectively. The circle and square are represented values of ${{\nu_{\rm min1}}}$ obtained from observed and model oscillation frequencies, respectively.}
\end{center}
\end{figure}

\subsection{Reference frequency comparison}
The minima obtained from the model and the observed oscillation frequencies are compared for $l$=0 because other $l$ values are included mixed modes.
 In Fig. 2, the $\Delta\nu$ versus $\nu$ 
graph of KIC 9776739 shows a 
harmony between the observed and model minimum oscillation frequencies.
As seen in the pattern formed by observed 
and model oscillation frequencies, are 
quite compatible with each other.

Fig. 5 plots the model and observed 
${{\nu_{\rm min0}}}$ and ${{\nu_{\rm min1}}}$ of each star 
in terms of plots of $\Delta\nu$.
Fig. 5 compares the values of observed and model ${{\nu_{\rm min0}}}$ and ${{\nu_{\rm min1}}}$. 
The asterisks and filled circles are represented 
values of ${{\nu_{\rm min0}}}$ determined by 
model and observed oscillation frequencies, respectively. 
The filled circles and squares are represented values 
of ${{\nu_{\rm min1}}}$ obtained from observed and model oscillation frequencies, respectively.
The figure clearly shows 
that the observed and model ${{\nu_{\rm min0}}}$ and ${{\nu_{\rm min1}}}$
are in excellent agreement.
This indicates that the observed and model reference frequencies 
are just as congruent as the observed and model values of $\Delta\nu$.
The ${{\nu_{\rm min0}}}$ and ${{\nu_{\rm min1}}}$ values
 are listed in Table 3.

In addition, the ${{\chi^2_{\rm seis}}}$ values are calculated from Eq. 5.
These values show that the model calculations 
for the stars are highly compatible with the observations.
The model and the observed oscillation frequencies in terms of 
${{\nu_{\rm min0}}}$, ${{\nu_{\rm min1}}}$, ${{\nu_{\rm max}}}$, and $\Delta\nu$
are used for calculated ${{\chi^2_{\rm seis}}}$. 
The calculated ${{\chi^2_{\rm seis}}}$ are listed in Table 3.

\begin{figure}
\begin{center}
\includegraphics[width=\columnwidth]{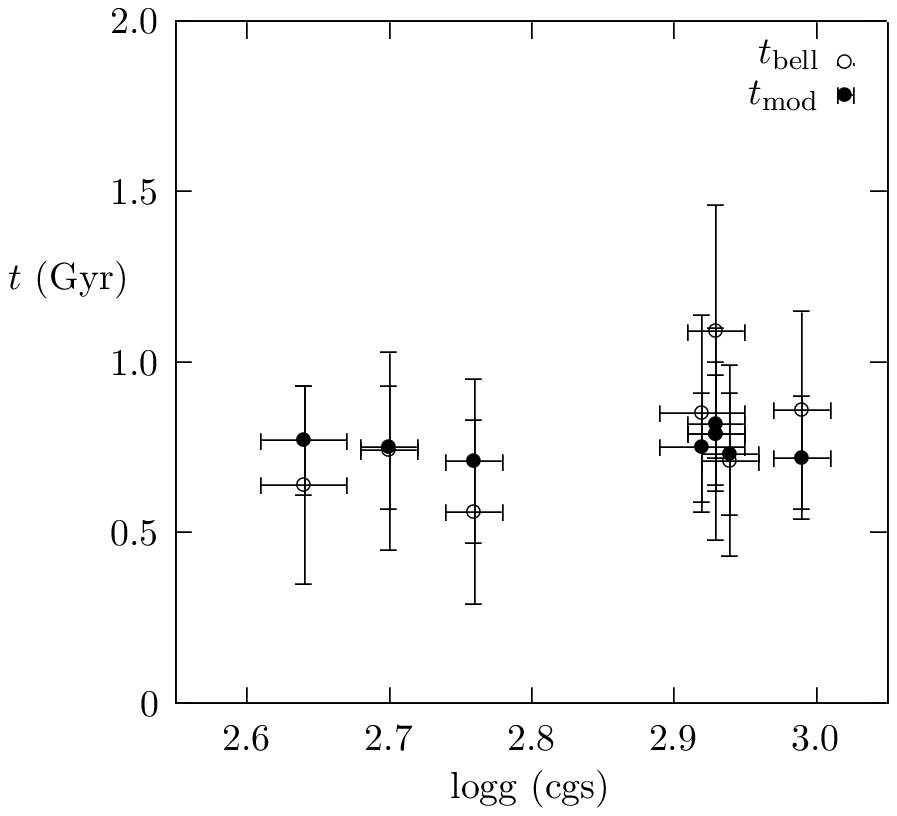}
        \caption{ The plot of stellar ages as function of the  $logg$  for the stars. The circles and filled circles are represented values of stellar age obtained from Bellinger (2019) and interior models, respectively. 
}
\end{center}
\end{figure}

\subsection{Age comparison}
Determining the stellar ages is critical for 
improve our understanding of stellar structure and evolution.
However, it is quite difficult to 
determine the stellar ages by observations.
Nevertheless, the age of a star is determined by interior models.
In this study, the stellar age is determined 
by constructed models using asteroseismic constraints.
Moreover, the solar-like oscillating 
stars studied are members of the open cluster. 
Thus, the range of model ages is  $0.71- 0.82$ Gyr.
The model stellar ages are listed in Table 2.

\begin{table}
\begin{center}
        \caption{The ages of the NGC 6811 obtained by different methods.
In the table, age and errors, and reference information are given, respectively. 
}
        \begin{tabular}{lcc}
                \hline
Age  &ref&\\ \\[1.0pt]

(Gyr)&&\\ \\[1.0pt]
  \hline
1.05$\pm$0.05& Sandquist et al. 2016\\[2.0pt]
1.04$\pm$0.07& Curtis et al. 2019\\[2.0pt]
1.00$\pm$0.02&Spada \& Lanzafame 2020\\[2.0pt]
1.00$\pm$0.07& Cargile et al. 2014\\[2.0pt]
1.00$\pm$0.02&Meibom et al. 2011\\[2.0pt]
1.00$\pm$0.04&Rodriguez et al. 2020\\[2.0pt] 
0.94$\pm$0.08& Janes et al. 2013\\[2.0pt]
0.87$\pm$0.02&Bossini et al. 2019\\[2.0pt]
0.70$\pm$0.10& Glushkova et al. 1999\\[2.0pt]
0.64$\pm$0.02& Donor et al. 2020\\[2.0pt]
               \hline
        \end{tabular}
\end{center}
\end{table}

The ages of NGC 6811 reported in different studies are listed in Table 5.
In Table 5, the oldest $t$ is 1.05 Gyr (Sandquist et al. 2016), while 
the youngest $t$ is 0.64 Gyr (Donor et al. 2020).
A reason accounting for these age differences could
 be the methods used.
For example, Meibom et al. (2011) 
examined the NGC 6811 cluster in detail, 
along with the relationship between the 
stellar rotation periods and their masses and ages. 
Thus, the age of this cluster was determined to be approximately 1 Gyr.
Another reason accounting for the difference 
may be model input chemical composition.
For example, 36 {\emph{Kepler}} 
short-cadence stars in 
NGC 6811 were anaysed by Rodriguez et al. (2020). 
Their star types include two solar-like oscillating red giant stars, 21 MS pulsators (16 $\delta$ Scuti and five $\gamma$ Doradus stars), 
and 13 rotating variables. 
They found that the age of cluster is 1 Gyr from Padova stellar isochrone 
for Z=0.012.
On the other hand, NGC 6811 cluster  is characterized as a young 
(0.7$\pm$0.1 Gyr) and possibly 
solar-metallicity star population by Glushkova et al. (1999).
In this study, the models constructed with 
the {\small {MESA}} evolution code are taken as $Z=0.0126$
and average age is approximately 0.76 Gyr.

Finally, the stellar ages are computed using the 
Python code developed by Bellinger (2019).
With this code, the ages of evolved stars 
are obtained by using observational 
asteroseismic parameters without constructed interior model.
The ages obtained from the model and Bellinger's method are compared in Fig. 6.
The $t$ range calculated from Bellinger is $0.56-0.74$ Gyr.
This age range is 
very compatible with the age range 
obtained by a model constructed in this study.
When the $t$ of the literature and the cluster members in this study are compared, 
the minimum difference is 0.01$\pm$0.28 Gyr.
This calculated value is still within the error range of the model ages.
Thus, it can be concluded that the different stellar ages obtained for 
these cluster members are quite compatible with the model ages.

\section{Conclusions}
This study analyses 
eight solar-like oscillating 
evolved stars which are NGC 6811 cluster members.
These selected stars are modelled for the first time in this study.
The stellar models have been constructed 
by using {\small {MESA}} code.
The interior models of the 
 stars are included the  mass-loss method.
The oscillation frequencies of eight evolved stars are examined. 
The mean large separation, frequencies of 
the maximum amplitude, and
reference frequencies are computed by stellar models.

The fundamental stellar properties 
 are determined in the interior models using 
these asteroseismic parameters.
The observed asteroseismic 
($\Delta\nu$, ${\nu_{\rm max}}$,  ${\nu_{\rm min0}}$, and ${\nu_{\rm min1}}$), and non-asteroseismic ($T_{\rm eff}$, $log g$, and [Fe/H]) parameters are compared with model values, which are obtained from {\small {MESA}} evolution 
code for the stars.
The compatibility between the model and
observed oscillation frequencies was examined.
For this, reference frequencies
and $\Delta\nu$ obtained from the model and observation are compared.
It was observed that values of the model 
${\nu_{\rm min1}}$ and ${\nu_{\rm min0}}$ are
compatible with the observational ${\nu_{\rm min1}}$ and ${\nu_{\rm min0}}$ frequencies.
Moreover, attention was paid to ensure that the model $\Delta\nu$ value is very close the observational value.
These methods allow us to 
determine the fundamental
 stellar parameters to high precision. 
The stellar
models are constructed using mass-loss methods ($\eta$=0.1).
The total mass loss in all the interior
models constructed is 0.01 $M_{\rm \odot}$.
The stars have masses and radii spanning of  
$2.23-2.40$ $M_{\rm \odot}$ and $8.47-12.38$ $R_{\rm \odot}$, 
respectively, with typical uncertainties of $\sim$ 11 per cent in mass and $\sim$ 9 per cent in radius.

Furthermore, the conventional, modified scaling relations, 
and the classical method have been used to calculated 
the masses and radii of the evolved stars.
The masses and radii obtained 
from the different methods are compared with those derived by the model.
The radii obtained from the 
model and those obtained from 
different methods are close to each other. 
However, the stellar mass determined by scaling relations, 
and the model mass is slightly different from each other.
In addition, the agreement between the model masses
and those  obtained from the classical method is very good at such stars.

After stellar mass, chemical composition is the most important 
parameter 
that has an impact on the stellar structure and evolution.
However, determining the metal abundance of a star is very difficult.
Studying cluster member stars with the 
same metal abundance provides an advantage in this regard.
The method widely reported in the
literature is the calculation of the
metal abundance of stars from [Fe/H] or [M/H].
In this study, it is calculated by averaging 
the calculated $Z$ values of the stars that are 
members of the NGC 6811 cluster.
The stellar metallicity
is calculated from mean metallicity and 
$Z$  is set to 0.0126$\pm$0.0003.

In this study, the stellar age is calculated 
from {\small {MESA}} models for the stars. 
Model stellar ages are listed in Table 2.
The stellar model age range is 0.71-0.82 Gyr.
In addition, the age determined using different methods 
was compared with the model ages.
When comparing the ages obtained from these different 
methods with {\small {MESA}} models,
 the minimum difference is 0.01$\pm$0.28 Gyr.
Also, the $t$ range calculated from the Python code developed 
by Bellinger (2019) is $0.56-0.74$ Gyr. 
The stellar $t$ obtained from the models is 
compatible with the literature.
The fact that the age values calculated by different 
methods are close to each other indicates how well the model ages are obtained with the methods applied in this study. 

\section*{Acknowledgements}
I would like to thank 
Dr. Mutlu Y{\i}ld{\i}z for sharing his 
invaluable experiences and contributing 
in this study with \textit{Gaia} database analysis, 
and Sibel \"{O}rtel, who has been a great help in modeling process.
I am grateful to Ege University Planning 
and Monitoring Coordination of Organizational 
Development and Directorate of Library and Documentation for their support in editing and proofreading service of this study. 
I would also like to 
thank my son Emirhan Sami and
my husband Mustafa Orhan for their support. 
This work is supported by the Scientific and Technological Research Council of Turkey (T\"{U}B\.{I}TAK:118F352).
I would like to thank Kelly Spencer for her kind help in checking the language of the revised manuscript.

~\\
{\uppercase{\bf{Data availability}}} \\
~\\
{The data underlying this article will be shared on reasonable request to the corresponding author.
} \\




\begin{thebibliography}{99}
\bibitem[\protect\citeauthoryear{Angelou}{1999}]{Angelou} { Angelou G. C., Bellinger E. P., Hekker S., Mints A., Elsworth Y., Basu S., Weiss A., 2020, MNRAS, 493, 4987}
\bibitem[\protect\citeauthoryear{Angulo}{1999}]{Angulo1999} Angulo C. et al., 1999, Nucl. Phys. A, 656, 3
\bibitem[\protect\citeauthoryear{Arentoft et al.}{2017}]{2017ApJ...838..115A} Arentoft T., Brogaard K., Jessen-Hansen J., Silva Aguirre V., Kjeldsen H., Mosumgaard J.~R., Sandquist E.~L., 2017, ApJ, 838, 115. doi:10.3847/1538-4357/aa63f7
\bibitem[\protect\citeauthoryear{Basu et al.}{2011}]{2011ApJ...729L..10B} Basu S., Grundahl F., Stello D., Kallinger T., Hekker S., Mosser B., Garc{\'\i}a R.~A., et al., 2011, ApJL, 729, L10. doi:10.1088/2041-8205/729/1/L10
\bibitem[\protect\citeauthoryear{Asplund et al.}{2009}]{2009ARA&A..47..481A} Asplund M., Grevesse N., Sauval A.~J., Scott P., 2009, ARA\&A, 47, 481. doi:10.1146/annurev.astro.46.060407.145222
\bibitem[\protect\citeauthoryear{Bedding et al.}{2011}]{2011Natur.471..608B} Bedding T.~R., Mosser B., Huber D., Montalb{\'a}n J., Beck P., Christensen-Dalsgaard J., Elsworth Y.~P., et al., 2011, Natur, 471, 608. doi:10.1038/nature09935
\bibitem[\protect\citeauthoryear{Bossini et al.}{2019}]{2019AA...623A.108B} Bossini D., Vallenari A., Bragaglia A., Cantat-Gaudin T., Sordo R., Balaguer-N{\'u}{\~n}ez L., Jordi C., et al., 2019, A\&A, 623, A108. doi:10.1051/0004-6361/201834693
\bibitem[\protect\citeauthoryear{Vitense}{1958}]{Vitense1958} B\"{o}hm- Vitense E., 1958, Z. Astrophys., 46, 108
\bibitem[\protect\citeauthoryear{Caughlan}{1988}]{Caughlan} Caughlan G. R., Fowler W. A., 1988, At. Data Nucl. Data Tables, 40, 283
\bibitem[\protect\citeauthoryear{Cargile et al.}{2014}]{2014ApJ...782...29C} Cargile P.~A., James D.~J., Pepper J., Kuhn R.~B., Siverd R., Stassun K.~G., 2014, ApJ, 782, 29. doi:10.1088/0004-637X/782/1/29
\bibitem[\protect\citeauthoryear{{\c{C}}elik Orhan, Y{\i}ld{\i}z, \& Kayhan}{2021}]{2021MNRAS.503.4529C} {\c{C}}elik Orhan Z., Y{\i}ld{\i}z M., Kayhan C., 2021, MNRAS, 503, 4529. doi:10.1093/mnras/stab757
\bibitem[\protect\citeauthoryear{Chaplin}{2014}]{Chaplin} Chaplin W. J. et al., 2014, ApJS, 210, 1
\bibitem[\protect\citeauthoryear{Christensen}{2008}]{chrish2008} Christensen-Dalsgaard J., 2008, Ap\&SS, 316, 113
\bibitem[\protect\citeauthoryear{Choi et al.}{2016}]{2016ApJ...823..102C} Choi J., Dotter A., Conroy C., Cantiello M., Paxton B., Johnson B.~D., 2016, ApJ, 823, 102. doi:10.3847/0004-637X/823/2/102
\bibitem[\protect\citeauthoryear{Curtis et al.}{2019}]{2019ApJ...879...49C} Curtis J.~L., Ag{\"u}eros M.~A., Douglas S.~T., Meibom S., 2019, ApJ, 879, 49. doi:10.3847/1538-4357/ab2393
\bibitem[\protect\citeauthoryear{Donor et al.}{2020}]{2020AJ....159..199D} Donor J., Frinchaboy P.~M., Cunha K., O'Connell J.~E., Allende Prieto C., Almeida A., Anders F., et al., 2020, AJ, 159, 199. doi:10.3847/1538-3881/ab77bc
\bibitem[\protect\citeauthoryear{Dotter}{2016}]{2016ApJS..222....8D} Dotter A., 2016, ApJS, 222, 8. doi:10.3847/0067-0049/222/1/8
\bibitem[\protect\citeauthoryear{Bellinger}{2019}]{Bellinger} Bellinger E. P., 2019, MNRAS, 486, 4612
\bibitem[\protect\citeauthoryear{Edvardsson}{1993}]{Edvardsson1993} Edvardsson, B., Andersen, J., Gustafsson, B, et al., 1993, A\&A,275, 101E
\bibitem[\protect\citeauthoryear{Ferguson}{2005}]{Ferguson2005} Ferguson J. W.,
Alexander D. R., Allard F., Barman T., Bodnarik J. G., Hauschildt P. h., Heffner- Wong A., Tammanai A., 2005, ApJ, 623, 585
\bibitem[\protect\citeauthoryear{Glushkova, Batyrshinova, \& Ibragimov}{1999}]{1999AstL...25...86G} Glushkova E.~V., Batyrshinova V.~M., Ibragimov M.~A., 1999, AstL, 25, 86
\bibitem[\protect\citeauthoryear{Guggenberger et al.}{2016}]{guggenber} Guggenberger E., Hekker S., Basu S., Bellinger E., 2016, MNRAS, 460,
4277
\bibitem[\protect\citeauthoryear{Hawkins et al.}{2016}]{2016A&A...594A..43H} Hawkins K., Masseron T., Jofr{\'e} P., Gilmore G., Elsworth Y., Hekker S., 2016, A\&A, 594, A43. doi:10.1051/0004-6361/201628812
\bibitem[\protect\citeauthoryear{Hekker et al.}{2011}]{2011AA...530A.100H} Hekker S., Basu S., Stello D., Kallinger T., Grundahl F., Mathur S., Garc{\'\i}a R.~A., et al., 2011, A\&A, 530, A100. doi:10.1051/0004-6361/201016303
\bibitem[\protect\citeauthoryear{Huang et al.}{2021}]{2021ApJ...910L...5H} Huang Y., Yuan H., Beers T.~C., Zhang H., 2021, ApJL, 910, L5. doi:10.3847/2041-8213/abe69a
\bibitem[\protect\citeauthoryear{Iglesis}{1993}]{igleses} Iglesias C. A., Rogers F. J., 1993, ApJ, 412, 752
\bibitem[\protect\citeauthoryear{Iglesis}{1996}]{igleses} Iglesias C. A., Rogers F. J., 1996, ApJ, 464, 943
\bibitem[\protect\citeauthoryear{Janes et al.}{2013}]{2013AJ....145....7J} Janes K., Barnes S.~A., Meibom S., Hoq S., 2013, AJ, 145, 7. doi:10.1088/0004-6256/145/1/7
\bibitem[\protect\citeauthoryear{Kjeldsen}{1995}]{kjelsen} Kjeldsen H., Bedding T. R., 1995, A\&A, 293, 87
\bibitem[\protect\citeauthoryear{Meibom et al.}{2011}]{2011ApJ...733L...9M} Meibom S., Barnes S.~A., Latham D.~W., Batalha N., Borucki W.~J., Koch D.~G., Basri G., et al., 2011, ApJL, 733, L9. doi:10.1088/2041-8205/733/1/L9
\bibitem[\protect\citeauthoryear{Miglio et al.}{2012}]{2012MNRAS.419.2077M} Miglio A., Brogaard K., Stello D., Chaplin W.~J., D'Antona F., Montalb{\'a}n J., Basu S., et al., 2012, MNRAS, 419, 2077. doi:10.1111/j.1365-2966.2011.19859.x
\bibitem[\protect\citeauthoryear{Mosser et al.}{2013}]{mosser13}Mosser B. et al., 2013, A\&A, 559, A137
\bibitem[\protect\citeauthoryear{Paxton}{2011}]{paxton2011} Paxton B., Bilsten L., Dotter A., Herwing F., Lesaffre P., Timmes F., 2011, ApJS, 2011, 192
\bibitem[\protect\citeauthoryear{Paxton}{2013}]{paxton2013} Paxton B., Catiello M., Arras P., Bildsten L., Brown E. F., Dotter A., Mankovich C., Montgomery M. H. et al., 2013, ApJS, 208
\bibitem[\protect\citeauthoryear{Reimers}{1975}]{1975MSRSL...8..369R} Reimers D., 1975, MSRSL, 8, 369
\bibitem[\protect\citeauthoryear{Renzini \& Fusi Pecci}{1988}]{1988ARAA..26..199R} Renzini A., Fusi Pecci F., 1988, ARA\&A, 26, 199. doi:10.1146/annurev.aa.26.090188.001215
\bibitem[\protect\citeauthoryear{Rodr{\'\i}guez et al.}{2020}]{2020MNRAS.491.4345R} Rodr{\'\i}guez E., Balona L.~A., L{\'o}pez-Gonz{\'a}lez M.~J., Ocando S., Mart{\'\i}n-Ruiz S., Rodr{\'\i}guez-L{\'o}pez C., 2020, MNRAS, 491, 4345. doi:10.1093/mnras/stz3143

\bibitem[\protect\citeauthoryear{Sandquist et al.}{2016}]{2016ApJ...831...11S} Sandquist E.~L., Jessen-Hansen J., Shetrone M.~D., Brogaard K., Meibom S., Leitner M., Stello D., et al., 2016, ApJ, 831, 11. doi:10.3847/0004-637X/831/1/11

\bibitem[\protect\citeauthoryear{Sharma al.}{2016}]{sharma} Sharma S., Stello D., Bland-Hawthorn J., Huber D., Bedding T. R., 2016,
ApJ, 822, 15
\bibitem[\protect\citeauthoryear{Spada \& Lanzafame}{2020}]{2020AA...636A..76S} Spada F., Lanzafame A.~C., 2020, A\&A, 636, A76. doi:10.1051/0004-6361/201936384
\bibitem[\protect\citeauthoryear{Stello et al.}{2010}]{2010AN....331..985S} Stello D., Basu S., Bedding T.\~R., Brogaard K., Bruntt H., Chaplin W.\~J., Christensen-Dalsgaard J., et al., 2010, AN, 331, 985. doi:10.1002/asna.201011442
\bibitem[\protect\citeauthoryear{Stello et al.}{2011}]{2011ApJ...739...13S} Stello D., Meibom S., Gilliland R.~L., Grundahl F., Hekker S., Mosser B., Kallinger T., et al., 2011a, ApJ, 739, 13. doi:10.1088/0004-637X/739/1/13
\bibitem[\protect\citeauthoryear{Stello et al.}{2011}]{2011ApJ...737L..10S} Stello D., Huber D., Kallinger T., Basu S., Mosser B.
, Hekker S., Mathur S., et al., 2011b, ApJL, 737, L10. doi:10.1088/2041-8205/737/1/L10
\bibitem[\protect\citeauthoryear{White}{2011}]{white} White T. R. et al., 2011, ApJ, 742, L3
\bibitem[\protect\citeauthoryear{Verma et al.}{2014}]{2014ApJ...790..138V} Verma K., Faria J.~P., Antia H.~M., Basu S., Mazumdar A., Monteiro M.~J.~P.~F.~G., Appourchaux T., et al., 2014, ApJ, 790, 138. doi:10.1088/0004-637X/790/2
\bibitem[\protect\citeauthoryear{Verma et al.}{2017}]{2017ApJ...837...47V} Verma K., Raodeo K., Antia H.~M., Mazumdar A., Basu S., Lund M.~N., Silva Aguirre V., 2017, ApJ, 837, 47. doi:10.3847/1538-4357/aa5da7
\bibitem[\protect\citeauthoryear{PaperI}{2014a}]{paperI} Y{\i}ld{\i}z M.,  \c{C}elik Orhan Z., Aksoy  C., Ok S., 2014a, MNRAS, 441, 2148 (Paper I)
\bibitem[\protect\citeauthoryear{yildiz2014b}{2014b}]{yildiz2014} Y{\i}ld{\i}z M.,  \c{C}elik Orhan Z., Kayhan  C., Turkoglu G. E., 2014b, MNRAS, 445, 4395
\bibitem[\protect\citeauthoryear{PaperII}{2015}]{paperII} Y{\i}ld{\i}z M.,  \c{C}elik Orhan Z., Kayhan C., 2015, MNRAS, 448, 3689 (Paper II)
\bibitem[\protect\citeauthoryear{PaperIII}{2016}]{paperIII} Y{\i}ld{\i}z M.,  \c{C}elik Orhan Z., Kayhan C., 2016, MNRAS, 462, 1577 (Paper III)



4277
\end{thebibliography}




\appendix

\section{Some extra material}

\begin{figure}
\begin{center}
\includegraphics[width=\columnwidth]{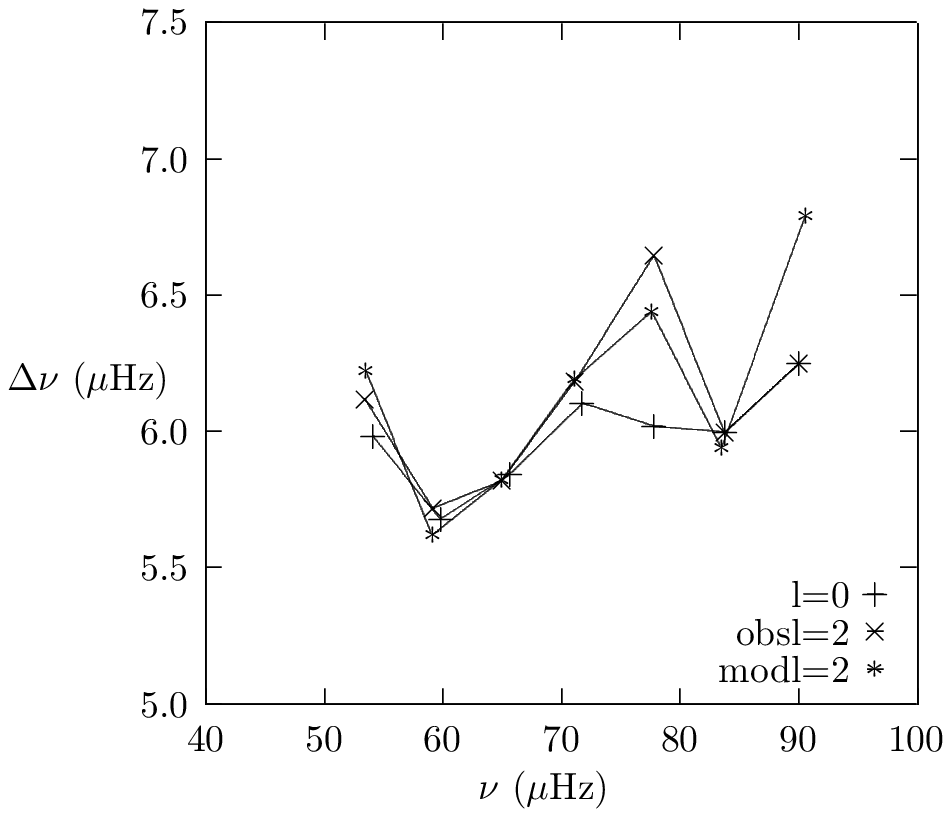}
        \caption{Plot of ${\Delta\nu}$ as a function of ${\nu}$ graph for $l$=2 model ({\small MESA}, asteriks) and observed $l$=0 and 2 oscillation frequencies (pluses and crosses, respectively) of KIC 9409513.
}
\end{center}
\end{figure}
If you want to present additional material which would interrupt the flow of the main paper,
it can be placed in an Appendix which appears after the list of references.


\bsp	
\label{lastpage}
\end{document}